\documentclass{ws-procs9x6}            

\usepackage{epsfig,amssymb,amsfonts,amsmath,mathtools,bm,color,xcolor,graphicx,braket,adjustbox,esint,upgreek}

\newcommand{\bonn}{Helmholtz-Institut f\"ur Strahlen- und Kernphysik and Bethe Center for Theoretical Physics, Universit\"at Bonn, D-53115 Bonn, Germany}

\newcommand{\fzj}{Institute for Advanced Simulation, Institut f\"ur Kernphysik and J\"ulich Center for Hadron Physics, Forschungszentrum J\"ulich, D-52425 J\"ulich, Germany}

\newcommand{\itep}{Institute for Theoretical and Experimental Physics, B. Cheremushkinskaya 25, 117218 Moscow, Russia}

\newcommand{\lebedev}{P.N. Lebedev Physical Institute of the Russian Academy of Sciences, 119991, Leninskiy Prospect 53, Moscow, Russia}

\newcommand{\scnu}{Institute of Quantum Matter, South China Normal University, Guangzhou 510006, China}

\newcommand{\tpcsf}{Theoretical Physics Center for Science Facilities, Institute of High Energy Physics, Chinese Academy of Sciences, Beijing 100049, China}

\newcommand{\ruhr}{Ruhr University Bochum, Faculty of Physics and Astronomy,\\ Institute for Theoretical Physics II, D-44780 Bochum, Germany }

\newcommand{\mephI}{National Research Nuclear University MEPhI, 115409, Kashirskoe highway 31, Moscow, Russia\\ $^*$E-mail: qianwang@m.scnu.edu.cn}

\newcommand{\zb}{$Z_b(10610)$}

\newcommand{\zbp}{$Z_b(10650)$}

\newcommand{\be}{\begin{equation}}
\newcommand{\ee}{\end{equation}}
\newcommand{\bea}{\begin{eqnarray}}
\newcommand{\eea}{\end{eqnarray}}
\newcommand{\beas}{\begin{eqnarray*}}
\newcommand{\eeas}{\end{eqnarray*}}

\graphicspath{{Figs.dir/}}

\begin{document}
\title{Implications of spin symmetry for $XYZ$ states}

\author{Q. Wang$^{1,2,*}$, V. Baru$^{3,4,5}$, E. Epelbaum$^{6}$, A. A. Filin$^6$, C. Hanhart$^7$, \\ A.V. Nefediev$^{5,8}$, J.L. Wynen$^7$}
\address{$^1$\scnu,\\ $^2$\tpcsf, \\ $^3$\bonn,\\ $^4$ \itep\\ $^5$ \lebedev,\\ $^6$\ruhr,\\$^7$\fzj,\\ $^8$\mephI,}


\begin{abstract}
Numerous exotic candidates containing a heavy quark and anti-quark (the so-called $XY\!Z$ states)
 have been reported since the observation of  the $X(3872)$ in 2003.
For these systems a study of the implications of the heavy quark spin symmetry and its breaking  is expected
 to provide useful guidance towards a better understanding of their nature.
 For instance, since the formation of the complete spin multiplets 
 is described with the same parameter sets, in some cases
the currently available experimental data on the $XY\!Z$ states allows us
to predict properties of spin partner states. To illustrate this point
 we extract the parameters of the two $Z_b$
states by analyzing the most recent experimental data within an effective-field theory approach
which treats both short-ranged contact interactions
and the long-ranged one-pion/one-eta Goldstone boson exchanges (OPE/OEE)  dynamically.
The line shapes and pole positions of their spin partners are then predicted in a parameter-free way and
await to be tested by future experimental data.

\end{abstract}


\bodymatter

\section{Introduction}\label{aba:sec1}
Heavy-quark spin symmetry (HQSS) states that the strong interaction is invariant
under the rotation of a spin of a heavy quark when the heavy quark mass goes to
infinity.  Corrections scale as $\Lambda_{QCD}/M_Q$
with $\Lambda_{QCD}\sim 200~\mathrm{MeV}$ the intrinsic mass scale of QCD
and $M_Q$ the heavy quark mass. Two important consequences follow: 1) in the heavy quark limit
the heavy quark spin and the total angular momentum of the light degrees of freedom
are conserved  separately by the strong interactions.  2) A given heavy quark spin state should
have spin partners with the same light cloud, but different heavy quark spins.
It can be shown that the violations of the spin symmetry predictions differ strongly
depending on the assumed structure of the states~\cite{Cleven:2015era}.
 In this contribution, we will use the two $Z_b$ states and their spin partners $W_{bJ}$ ($J=0,1,2$)
as an example to illustrate how the program works within the hadronic molecular picture, where
the leading order spin symmetry violation enters through $B-B^*$ mass difference.
In particular, an effective field theory (EFT),
which incorporates both short-range and long-range OPE interactions at leading order, is proposed
  for the scattering processes involving $\left(B,B^*\right)$ and $\left(\bar{B},\bar{B}^*\right)$ doublets.
  The parameters are extracted from the data on the line shapes of the two $Z_b$
states and used to predict the line shapes of their spin partners $W_{bJ}$ as well as their pole positions.

\section{Framework}
As  the bottom quark mass is much larger than the typical QCD scale,
i.e.~$M_b\gg \Lambda_{QCD}$, the heavy quark symmetry should be accurate for
bottomonium-like systems, such as the \zb, \zbp~and their spin partners.
Both the contact, OPE and OEE potentials between the $\left(B,B^*\right)$ and $\left(\bar{B},\bar{B}^*\right)$ doublets
were included within the EFT framework, as formulated in Refs.~\cite{Baru:2019xnh,Wang:2018jlv}.
In addition, the inelastic channels, i.e.~the $\left( \Upsilon(nS)\pi, \eta_b(nS)\pi \right)$ and $\left(\chi_{bJ}(mP)\pi, h_b(mP)\pi\right)$ channels
with $n=1,2,3$ and $m=1,2$ are also considered dynamically to satisfy the unitarity requirements.
Inelastic contributions are taken into account via corrections to the potentials between the elastic channels,
as the direct transitions between the inelastic channels can be neglected~\cite{Baru:2019xnh,Wang:2018jlv,Guo:2016bjq,Hanhart:2015cua}.
As the charge parities of the two $Z_b$ and their spin partners are different,
the negative  C-parity states  $Z_b$ can be produced from the initial $\Upsilon(10860)$ through an emission of one pion  $\Upsilon(10860)\to\pi Z_b^{(\prime)}\to\pi B^{(*)}\bar{B}^{(*)}$
while their positive C-parity partners, $W_{bJ}$, require an emission of the photon
 $\Upsilon(10860)\to\gamma W_{bJ}\to\gamma B^{(*)}\bar{B}^{(*)}$.
With the production amplitudes obtained by solving
the partial-wave-decomposed coupled channel Lippmann-Schwinger equation numerically~\cite{Baru:2019xnh,Wang:2018jlv},
we   obtained the invariant mass distributions for the $Z_b$s and their spin partners.
By adjusting the unknown parameters to data in the $Z_b$ channels and employing  HQSS
we predict the  line shapes for the  $W_{bJ}$'s.

\section{Results and discussions}

\begin{figure*}[ht]
\begin{center}
\epsfig{file=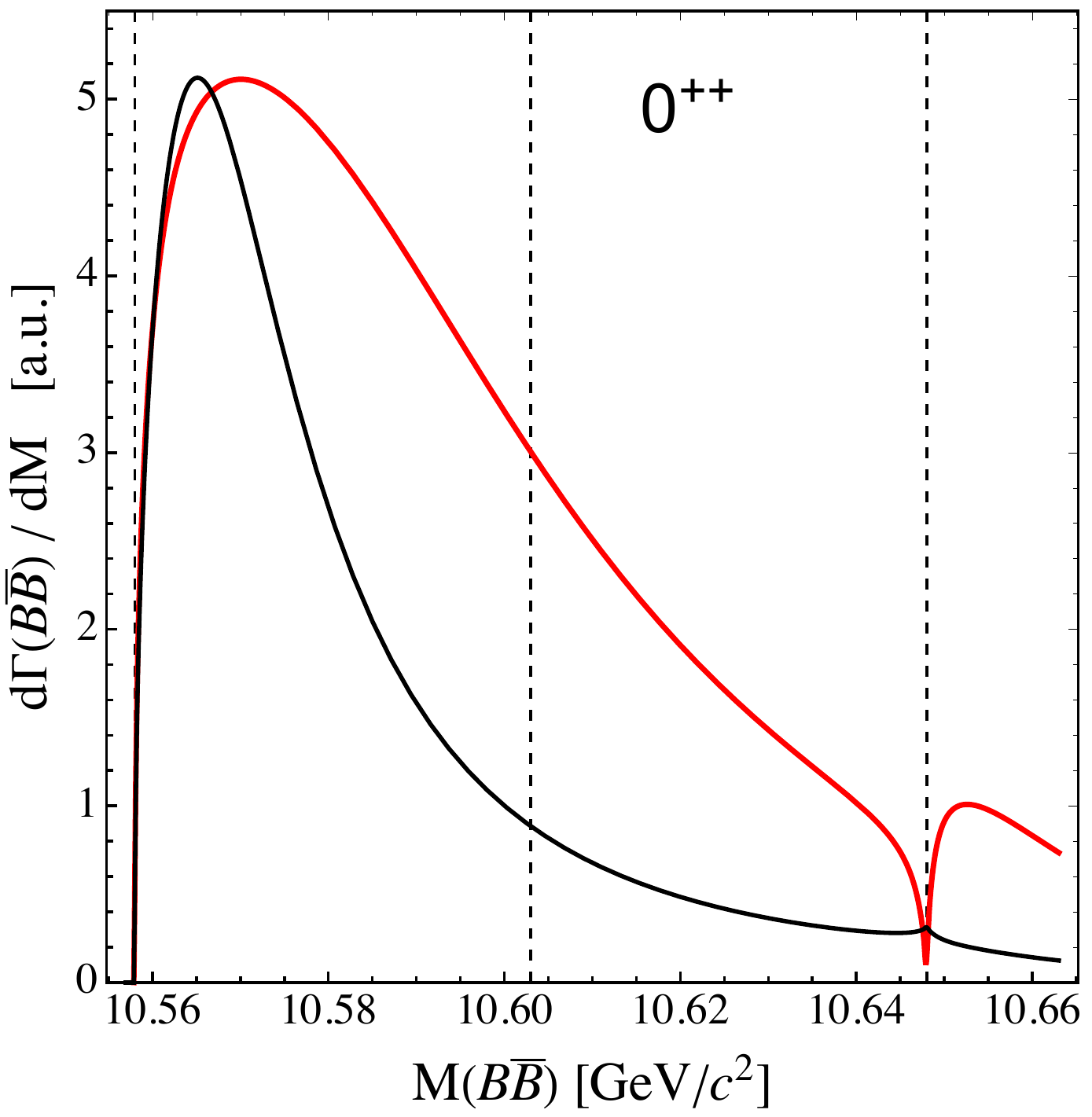, width=0.3\textwidth}
\epsfig{file=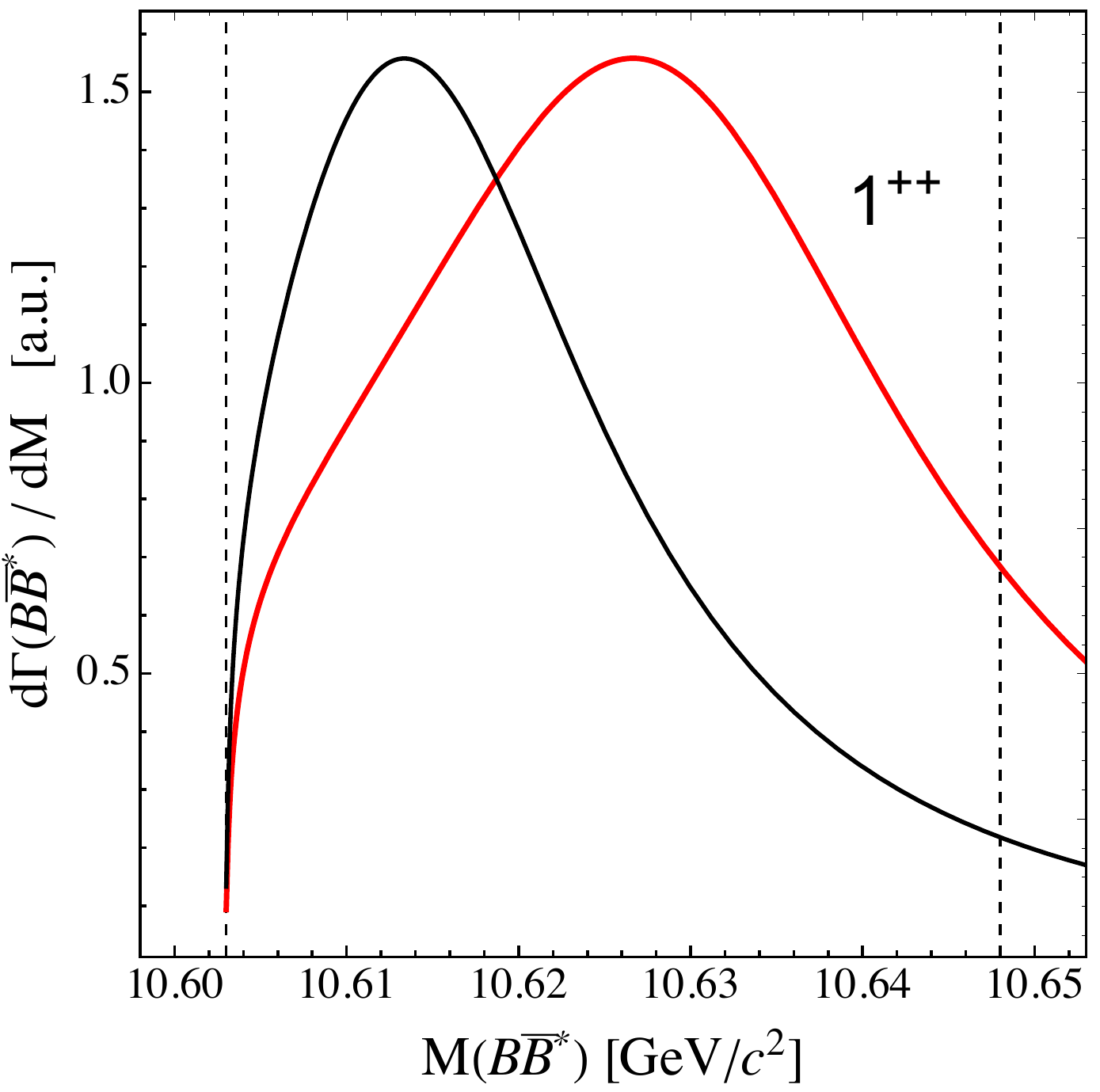, width=0.31\textwidth}
\epsfig{file=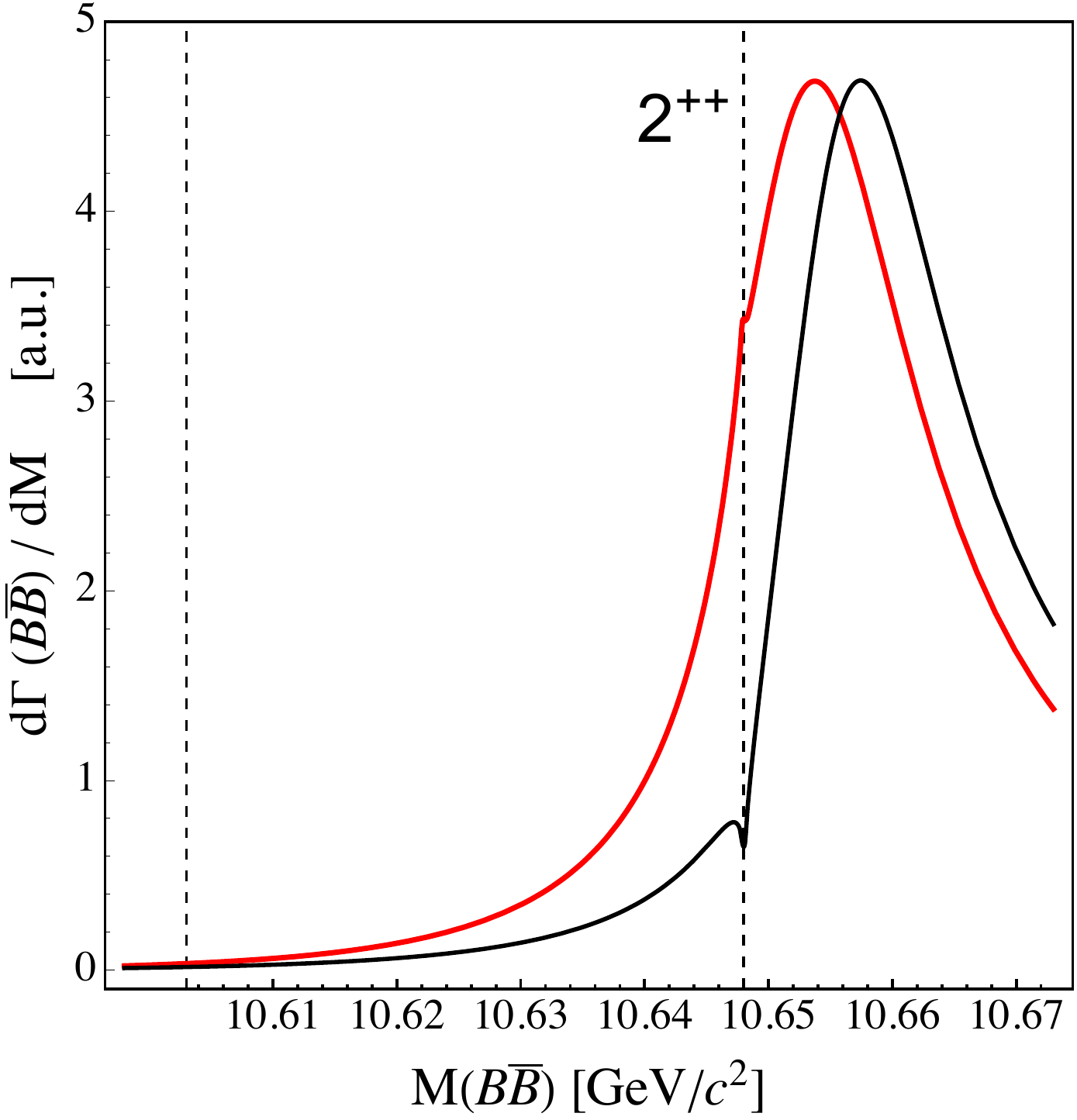, width=0.3\textwidth}
\epsfig{file=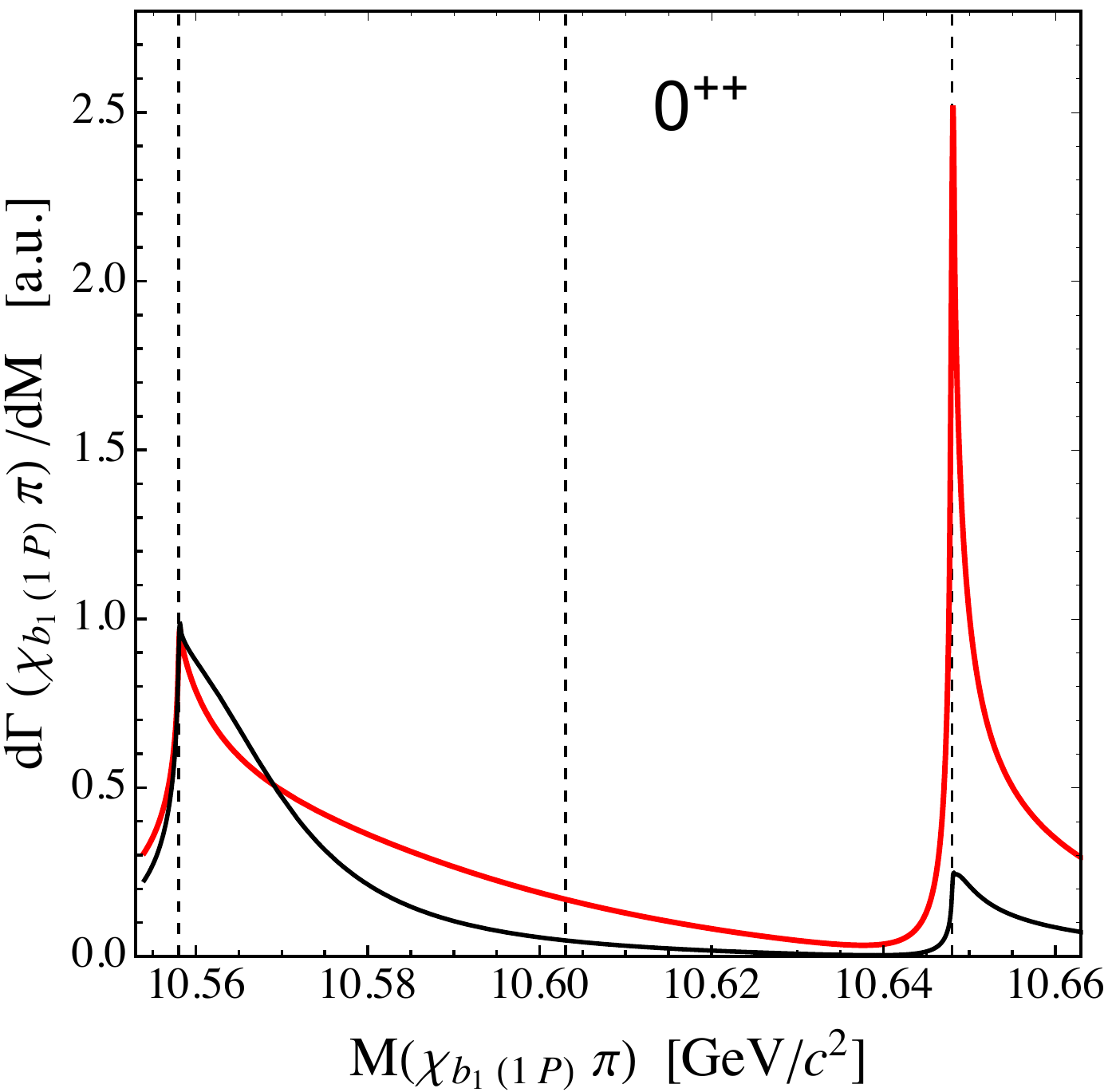, width=0.3\textwidth}
\epsfig{file=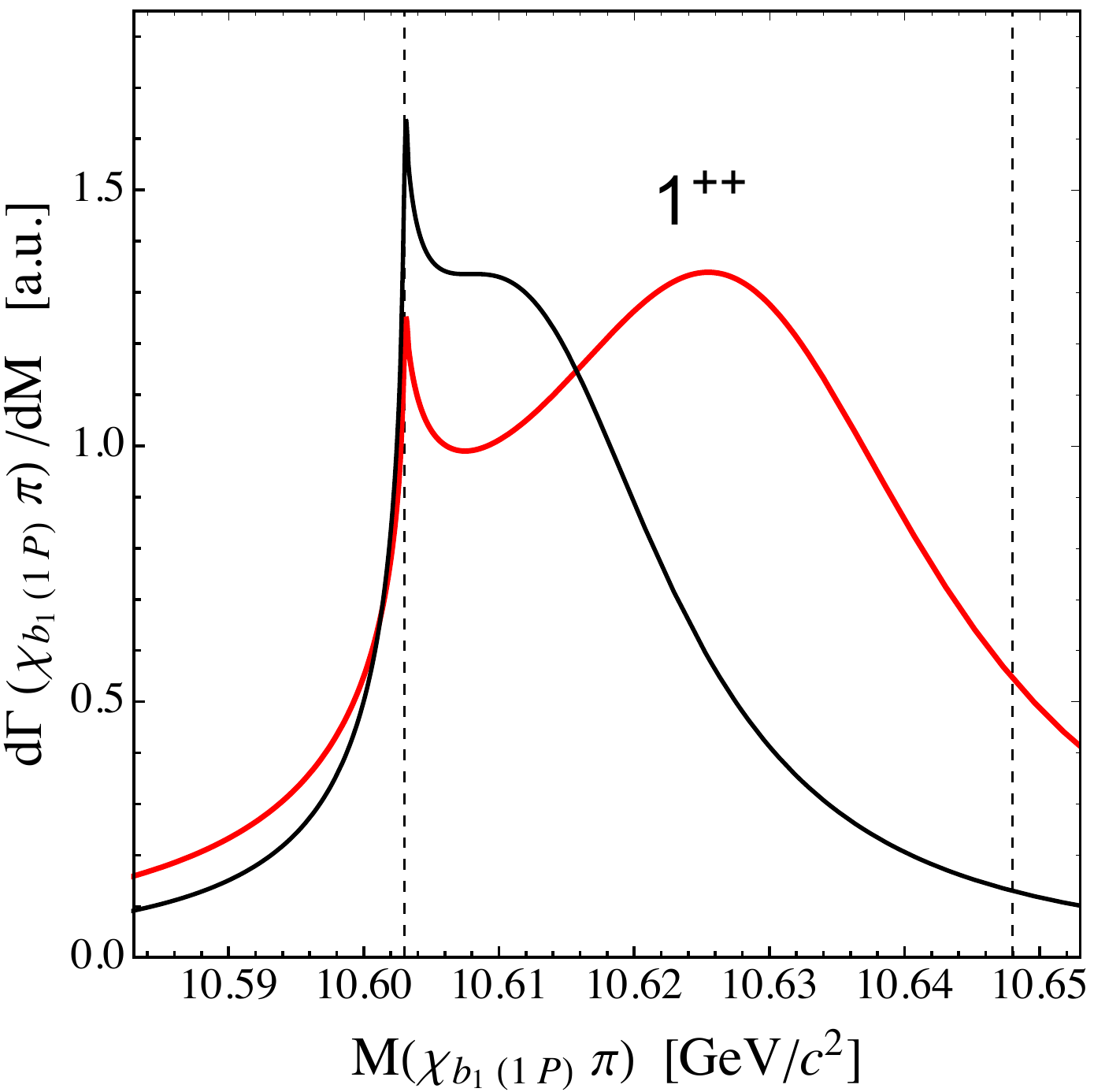, width=0.3\textwidth}
\epsfig{file=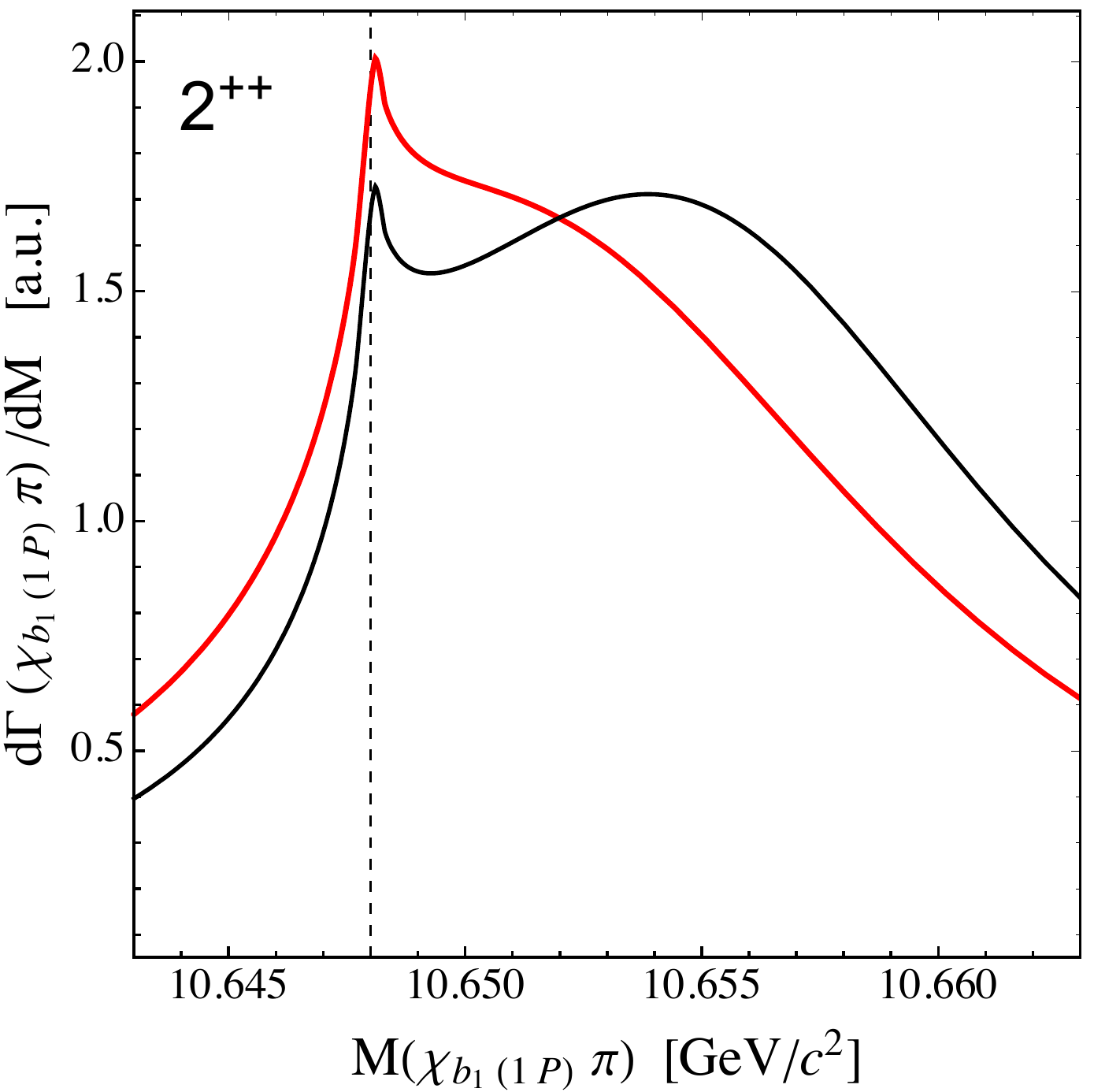, width=0.3\textwidth}
\caption{Predicted line shapes for the $0^{++}$, $1^{++}$ and $2^{++}$ channels in the corresponding
lowest elastic channels and the $\chi_{b1}(1P)\pi$ channels. Upper panel:
the line shapes of the $W_{b0}^{(\prime)}$, $W_{b1}$ and $W_{b2}$ in the $B\bar B$, $B\bar{B}^*$ and $B^*\bar B^*$ channels, respectively.
 Lower panel: the line shapes of the spin partners $W_{bJ}$ states in the $\chi_{b1}(1P)\pi$
 channel. The red and black lines show the results for schemes II and III, respectively, and the vertical dashed lines indicate the position of the $B\bar{B}, B\bar{B}^*$
and $B^*\bar{B}^*$ thresholds.}\label{fig:0pp}
\end{center}
\end{figure*}

\begin{center}
\begin{table}
\tbl{The pole positions  (on the sheet close to the physical one) and the residues in various $S$-wave $B^{(*)}\bar B^{(*)}$ channels for Scheme III.
The energy $E_{\rm pole}$ is given relative to the nearest open-bottom threshold quoted in the third column.
Uncertainties correspond to a $1\sigma$
deviation in the parameters allowed by the fit to the data in the $Z_b$s' channels. The poles are
calculated for the cutoff $\Lambda=1$ GeV. }
{\tablefont
\begin{tabular}{lllll}
\toprule $J^{PC}$ & State & Threshold & $E_{\mathrm{pole}}$ w.r.t. threshold $[\mathrm{MeV}]$ & Residue at $E_{\mathrm{pole}}$\\\colrule
$1^{+-}$ & $Z_b$ &  $B\bar{B}^*$&$(-2.3\pm 0.5)-i(1.1\pm 0.1)$ & $ (-1.2\pm 0.2)+i(0.3\pm 0.2) $\\
$1^{+-}$ & $Z_b^\prime$ & $B^*\bar{B}^*$ & $(1.8 \pm 2.0) - i (13.6 \pm 3.1)$ & $(1.5 \pm 0.2) - i (0.6 \pm 0.3)$\\
$0^{++}$ & $W_{b0}$ & $B\bar{B}$ &$(2.3 \pm 4.2) - i (16.0 \pm 2.6)$ & $(1.7 \pm 0.6) - i (1.7 \pm 0.5)$ \\
$0^{++}$ & $W_{b0}^\prime$ & $B^*\bar{B}^*$ &$(-1.3 \pm 0.4) - i (1.7 \pm 0.5)$ & $(-0.9 \pm 0.3) - i (0.3 \pm 0.2)$\\
$1^{++}$ & $W_{b1}$ & $B\bar{B}^*$ &$(10.2 \pm 2.5) - i (15.3 \pm 3.2)$ & $(1.3 \pm 0.2) - i (0.4 \pm 0.2)$\\
$2^{++}$ & $W_{b2}$ & $B^*\bar{B}^*$ &$(7.4 \pm 2.8) - i (9.9 \pm 2.2)$ & $(0.7 \pm 0.1) - i (0.3 \pm 0.1)$\\\botrule
\end{tabular}}\label{aba:theo}
\end{table}
\end{center}
In what follows, we present three fitting schemes, namely
 {\bf Scheme I}: purely $S$-wave momentum-independent contact interactions;
 {\bf Scheme II}: full OPE potential as well as the ${\cal O}(Q^2)$ $S$-wave-to-$D$-wave
 counter term promoted to LO in addition to scheme I;
 {\bf Scheme III}: the ${\cal O}(Q^2)$ $S$-wave-to-$S$-wave contact terms at NLO and the
$\eta$-meson exchange potential on top of scheme II.
\begin{figure*}[t]
\begin{center}
\epsfig{file=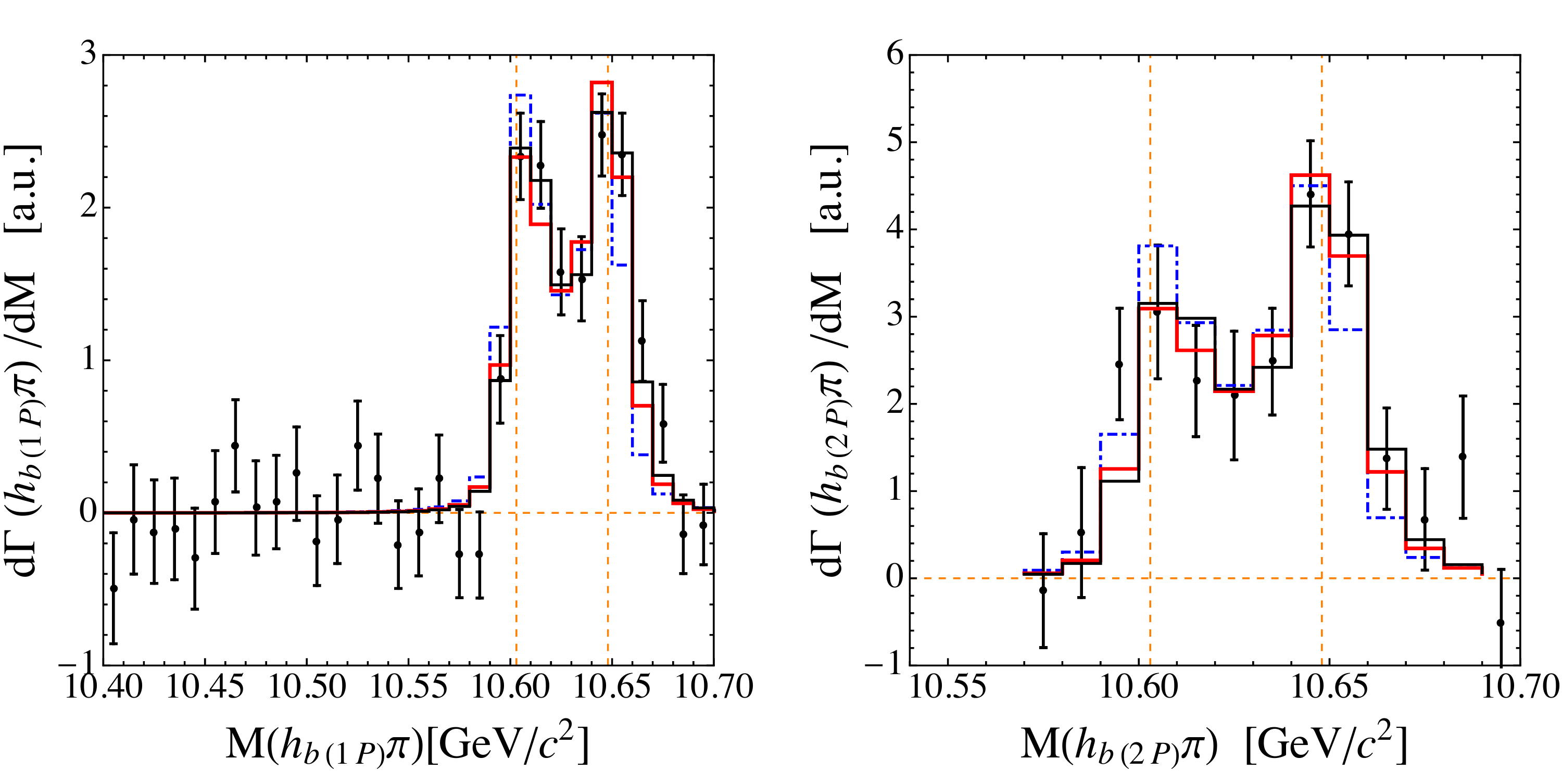, width=0.6\textwidth}\\
\caption{The fitted line shapes in the $1^{+-}$ channel in the
 $h_b(1P)\pi$ and $h_b(2P)\pi$ channels for an illustration~\cite{Baru:2019xnh}.
  The line shapes of schemes I, II and III are
shown by the blue dashed, red thick solid and black solid curves, respectively.
The vertical dashed lines indicate the positions of the $B\bar{B}^*$ and $B^*\bar{B}^*$ thresholds.
The experimental data are from Refs.~\cite{Belle:2011aa}}\label{fig:1pm}
\end{center}
\end{figure*}
The fit results of Scheme I agree with those of the parametrization proposed in Refs~\cite{Guo:2016bjq,Hanhart:2015cua}, which
demonstrates that  the two approaches are consistent with each other.
As expected, the inclusion of the pion dynamics in Scheme II visibly improves the fit
 once the ${\cal O}(Q^2)$ $S$-wave-to-$D$-wave
 counter term is considered simultaneously,
 which is reflected in the reduction of  $\chi^2/\mathrm{d.o.f.}$ from 1.29 to 0.95,   see also Fig.~\ref{fig:1pm}.
 The fit shows that HQSS violating  the contact interactions are not required by the data.
The transition to Scheme III represents a rather perturbative effect which leads to some further (although small) reduction of   $\chi^2/\mathrm{d.o.f.}$ to 0.83, however,
the inclusion of the
 ${\cal O}(Q^2)$ $S$-wave-to-$S$-wave contact terms 
 allows one to  remove some higher-order regulator artefacts related with  the iteration of the truncated potential within the integral equations
 (see Ref.~\cite{Baru:2019xnh} for the extended discussion).
Using Scheme III as our final result we
 predict the line shapes (Fig.~\ref{fig:0pp}) and the pole positions (Table~\ref{aba:theo}) of the $W_{bJ}$s in a parameter-free way.
Our predictions for the line shapes are strongly asymmetric (either a bump just above threshold or a large
 sizable distortion at threshold) reflecting the assumed molecular nature of the states. Thus, an experimental
 conformation of those predictions would provide strong support for the hadronic molecule picture for both groups
 the $W_b$-states as well as the $Z_b$ states.
\section{Summary}
A systematic EFT approach with respect to both chiral and heavy-quark spin symmetries
is proposed for the two $Z_b$ states. In our framework,
both the short-range contact and the long-range OPE/OEE potentials are considered dynamically,
as well as the inelastic channels.  With the parameters extracted from the invariant mass distributions of the two $Z_b$ states,
the line shapes and pole positions of the spin
partners $W_{bJ}$s are predicted in a parameter-free way.
They can be confronted with the results from
future high-luminosity and high-statistic experiments, such as Belle-II,
to gain insights into the nature of these exotic candidates.
\section{Acknowledgments}
This work is supported in part by the National Natural Science Foundation of China (NSFC) and  the
Deutsche Forschungsgemeinschaft (DFG) through the funds provided to the Sino-German Collaborative Research
Center ``Symmetries and the Emergence of Structure in QCD''  (NSFC Grant No. 11621131001 and DFG Grant No. TRR110).
QW is also supported by the
Thousand Talents Plan for Young Professionals and research startup funding at SCNU.
Work of V.B. and A.N. was supported by the Russian Science Foundation (Grant No. 18-12-00226).


\end{document}